%% file: pv-conf.tex
\documentclass{vgtc}                          

\ifpdf
  \pdfoutput=1\relax                   
  \usepackage{graphicx}                
  \DeclareGraphicsExtensions{.pdf,.png,.jpg,.jpeg} 
\else
  \usepackage{graphicx}                
  \DeclareGraphicsExtensions{.eps}     
\fi%

\graphicspath{{figures/}{pictures/}{images/}{./}} 

\usepackage{microtype}                 
\PassOptionsToPackage{warn}{textcomp}  
\usepackage{textcomp}                  
\usepackage{mathptmx}                  
\usepackage{times}                     
\usepackage{cite}                      
\usepackage{tabu}                      
\usepackage{booktabs}                  
\usepackage{amsmath}
\usepackage{amssymb}
\usepackage{amsfonts}
\usepackage{graphicx}
\usepackage{pifont}
\usepackage{gensymb}
\usepackage[normalem]{ulem}
\usepackage{bm}
\usepackage{cancel}
\usepackage{color}

\onlineid{0}

\vgtccategory{Research}

\vgtcinsertpkg

\title{Exact Analytical Parallel Vectors}

\author{Hanqi Guo
\and Tom Peterka\thanks{e-mails: \{hguo$\vert$tpeterka\}@anl.gov}}\affiliation{\scriptsize Mathematics and Computer Science Division, Argonne National Laboratory}

\abstract{This paper demonstrates that parallel vector curves are piecewise cubic rational curves in 3D piecewise linear vector fields.  Parallel vector curves---loci of points where two vector fields are parallel---have been widely used to extract features including ridges, valleys, and vortex core lines in scientific data.  We define the term \emph{generalized and underdetermined eigensystem} in the form of $\mathbf{A}\mathbf{x}+\mathbf{a}=\lambda(\mathbf{B}\mathbf{x}+\mathbf{b})$ in order to derive the piecewise rational representation of 3D parallel vector curves.  We discuss how singularities of the rationals lead to different types of intersections with tetrahedral cells.} 

\begin{document}


\firstsection{Introduction}
\maketitle

\input{introduction.tex}

\input{background.tex}
\input{math.tex}

\input{algorithm.tex}
\input{conclusions.tex}

\acknowledgments{We thank Drs. Chunhui Liu and Xin Liang for useful discussions.  This research is supported by the Exascale Computing Project (ECP), project number 17-SC-20-SC, a collaborative effort of the U.S. Department of Energy Office of Science and the National Nuclear Security Administration. 
It is also supported by the U.S. Department of Energy, Office of Advanced Scientific Computing Research, Scientific Discovery through Advanced Computing (SciDAC) program, and by Laboratory Directed Research and Development (LDRD) funding from Argonne National Laboratory, provided by the Director, Office of Science, of the U.S. Department of Energy under Contract No. DE-AC02-06CH11357.  This work is also supported in part by National Science Foundation Division of Information and Intelligent Systems-1955764.}

\bibliographystyle{abbrv-doi}

\bibliography{pv-conf}

\vfill
{\footnotesize The submitted manuscript has been created by UChicago Argonne, LLC, Operator of Argonne National Laboratory (``Argonne''). Argonne, a U.S. Department of Energy Office of Science laboratory, is operated under Contract No. DE-AC02-06CH11357. The U.S. Government retains for itself, and others acting on its behalf, a paid-up, nonexclusive, irrevocable worldwide license in said article to reproduce, prepare derivative works, distribute copies to the public, and perform publicly and display publicly, by or on behalf of the Government.The Department of Energy will provide public access to these results of federally sponsored research in accordance with the DOE Public Access Plan. http://energy.gov/downloads/doe-public-access-plan.}

\end{document}

%% file: introduction.tex
\firstsection{Introduction}
\maketitle

The extraction of many one-dimensional feature curves---ridges, valleys, and vortex core lines---can be boiled down to the \emph{parallel vector} problem~\cite{PeikertR99}:
\begin{equation} \label{eq:pv}
  \mathbf{v}(\mathbf{x}) \times \mathbf{w}(\mathbf{x}) = \mathbf{0}~\mbox{or}~\mathbf{v}=\lambda\mathbf{w}, 
\end{equation}
where $\mathbf{v}, \mathbf{w}: \mathbb{R}^3 \to \mathbb{R}^3$ are two 3D vector fields and $\lambda$ is a real number.  The solutions $\mathbf{x}$ are a locus of points that normally form one-dimensional curves embedded in the 3D space.  For example, the Sujudi--Haimes vortex core descriptor~\cite{SujudiH95} can be interpreted as $\mathbf{v}\times((\nabla\mathbf{v})\mathbf{v})=\mathbf{0}$, where $\mathbf{v}$ is velocity; the Bank-Singer vortex core can be defined as $(\nabla\times\mathbf{v})\times(\nabla p)=\mathbf{0}$, $p$ being the pressure field; and ridge and valley lines can be formulated as $\mathbf{g}\times((\nabla\mathbf{g}^\intercal)\mathbf{g})=\mathbf{0}$, where $\mathbf{g}$ is the gradient field of a scalar field.

Challenges of extracting parallel vector curves (or simply PV curves) include both specificity and accuracy.  First, specificity becomes a problem when parallel vector curves are too close to each other, causing ambiguities in reconstructing the topology of the curves.  For example, the seminal work by Peikert and Roth~\cite{PeikertR99} uses a numerical method to find intersections between PV curves and individual mesh cells; when more than two intersections are detected, heuristics have been used to pair the intersections.  
Second, the reconstruction of PV curves inside mesh cells, in  other words the ``subpixel'' accuracy, is challenging.  For example, differential-equation-based methods such as feature flow fields (FFFs) exist but are subject to integration errors~\cite{TheiselS03}.

In this study we present a mathematical derivation to extract analytical exact PV curves.  We regard our method as a generalization of the work of Peikert and Roth~\cite{PeikertR99}. We demonstrate a variety of uses of analytical PV curves and envision the future possibilities of using analytical PV curves.  Overall, the contribution of this paper is twofold: 
\begin{itemize}
  \item Theoretical contribution that PV curves are piecewise cubic rational curves in piecewise linear vector fields;
  \item An analytical exact PV curve extraction algorithm.
\end{itemize}

%% file: background.tex
\section{Background}
\label{sec:background}

This section first reviews the classical Peikert--Roth method and then discusses other approaches to extract PV curves.  




\subsection{Peikert-Roth method}


The Peikert--Roth method~\cite{PeikertR99} is a fundamental approach to extract PV curves in 3D vector fields.  The basic assumption is that $\mathbf{v}$ and $\mathbf{w}$ are linear on triangular faces (2-simplices) in the mesh; subdivision of non-triangular faces is needed if the mesh is nonsimplicial. Based on the linearity assumption, one can find intersections between PV curves and every triangular face by solving the following equation:
\begin{equation}\label{eq:eig3}
  \left(
  \begin{matrix}
    v_{0x} & v_{1x} & v_{2x}\\
    v_{0y} & v_{1y} & v_{2y}\\
    v_{0z} & v_{1z} & v_{2z}
  \end{matrix}
  \right)
  \left(
  \begin{matrix}
    \mu_0 \\
    \mu_1 \\
    \mu_2
  \end{matrix}
  \right) = \lambda
  \left(
  \begin{matrix}
    w_{0x} & w_{1x} & w_{2x}\\
    w_{0y} & w_{1y} & w_{2y}\\
    w_{0z} & w_{1z} & w_{2z}
  \end{matrix}
  \right)
  \left(
  \begin{matrix}
    \mu_0 \\
    \mu_1 \\
    \mu_2
  \end{matrix}
  \right), 
\end{equation}
where columns $(v_{ix}, v_{iy}, v_{iz})^\intercal$ and $(w_{ix}, w_{iy}, w_{iz})^\intercal$ denote the $xyz$ components of $\mathbf{v}$ and $\mathbf{v}$ on the $i$th node of the triangle; and $\bm{\mu}=(\mu_0, \mu_1, \mu_2)^T$ denotes the barycentric coordinates and $\mu_0+\mu_1+\mu_2=1$.  If the result barycentric coordinates are within $[0, 1]$, the intersection is in the triangle and is a \emph{parallel vector point} (PV point). 

Equation~\eqref{eq:eig3}, known as the \emph{generalized eigenvalue} problem in the form of $\mathbf{A}\bm{\mu}=\lambda\mathbf{B}\bm{\mu}$ ($\mathbf{A}$ and $\mathbf{B}$, respectively, represents the $3\times3$ matrix in the left- and right-hand side of the equation), has closed-form solutions of eigenvalues $\lambda$ and eigenvectors $\bm{\mu}$.  There exists a method to transform the equation into a \emph{characteristic polynomial}
\begin{equation}\label{eq:poly3}
\operatorname{det}(\mathbf{A})\lambda^3 + \cdot\lambda^2 + \cdot\lambda + \operatorname{det}(\mathbf{B}),
\end{equation}
where we omit, for now, the quadratic and linear coefficients with the dot ($\cdot$) symbol, for clarity.  

A limitation of this method is the specificity and accuracy when reconstructing PV curves from PV points.  First, one may associate two intersections if they are on the triangular faces of the same 3D cell, but ambiguity exists when the cell has more than two intersections.  Actually, each triangular face may have up to three PV points because the characteristic polynomial is cubic, and chances exist that a 3D cell has many intersections.  In such cases, heuristics or cell subdivision are needed to pair the intersections.
Second, although the PV points are analytically exact on 2D triangular faces, no such exact method exists for characterizing PV points and curves inside 3D cells.  Our method generalizes the  Peikert--Roth method and resolves both specificity and accuracy problems with an exact analytical solution, as described in the rest of this paper.


\subsection{Other methods}

Various methods are proposed to address the specificity and accuracy problem; to date, however, no method delivers exact analytical results.

A \textbf{parity test} method was proposed in ~\cite{JuCWD14} to eliminate ambiguities when multiple pairs of PV points exist on the faces of a cell.  Ambiguity cases produced by the Peikert--Roth method can be used as the input, and the parity test samples $\mathbf{u}$ and $\mathbf{v}$ on the boundary of the faces and uses Poincare--Hopf and Gauss--Bonnett theorems to determine the parity.  

An \textbf{isosurface-based method}~\cite{PeikertR99} views PV curves as the intersection between two isosurfaces---zero-level sets of the $x$- and $y$- components of $\mathbf{v}(\mathbf{x})\times\mathbf{w}(\mathbf{x})$.  This method assumes linearity of the cross product field and can be numerically challenging; one has to verify the $z$-component of the product is zero in the outputs.

\textbf{Integral-based method} such as feature flow fields (FFF)~\cite{TheiselS03} views PV curves as integral curves of a derived vector field, namely feature flow fields.  Although methods exist to improve stability of FFFs~\cite{WeinkaufTGP11}, error accumulates in solving ordinary differential equations (ODEs) and transforming input vector fields into FFFs.  In order to eliminate error accumulation in solving ODEs, Gelder and Pang~\cite{GelderP09} proposed PVSolve, which uses the dimensionless project vector at every iteration and enables larger step sizes than FFF methods.  In addition, integral-based methods have been generalized to high-order meshes~\cite{PagotOSWEC11} and time-tracking~\cite{TheiselSWHS05}.  






%% file: math.tex
\section{Mathematical Formulation}
\label{sec:math}

This section proves that PV curves in piecewise linear vector fields $\mathbf{v}(\mathbf{x})$ and $\mathbf{w}(\mathbf{x})$ are parametric curves and piecewise cubic rational functions of $\lambda$.  



\subsection{Assumption}

We assume that both $\mathbf{v}(\mathbf{x})$ and $\mathbf{w}(\mathbf{x})$ in Equation~\eqref{eq:pv} are piecewise linear (PL).  The PL assumption implies that the domain is discretized into 3D simplicial tetrahedral cells.  In each tetrahedron, both $\mathbf{v}(\mathbf{x})$ and $\mathbf{w}(\mathbf{x})$ can be linearly interpolated:
\begin{equation}
  \begin{array}{rl}
  \mathbf{v}(\mathbf{x})=& \mu_0\mathbf{v}_0 + \mu_1\mathbf{v}_1 + \mu_2\mathbf{v}_2 + \mu_3\mathbf{v}_2\\
  \mathbf{w}(\mathbf{x})=& \mu_0\mathbf{w}_0 + \mu_1\mathbf{w}_1 + \mu_2\mathbf{w}_2 + \mu_3\mathbf{w}_2\\
  1=& \mu_0 + \mu_1 + \mu_2 + \mu_3
  \end{array},
\end{equation}
where $(\mu_0, \mu_1, \mu_2, \mu_3)^\intercal$ are the \emph{barycentric coordinates} of a point on the PV curve; $\mathbf{v}_i$ and $\mathbf{w}_i$ ($i=0, 1, 2, 3$) are the vector values at the $i$th vertex of the tetrahedron.  We  consider only the PV curves inside the tetrahedron; that is, $\mu_i\in[0,1]$, $i=0, 1, 2, 3$.  

Based on the PL assumption, we rewrite Equation~\eqref{eq:pv} as 
\begin{equation}
  \left(\begin{matrix}
      v_{0x} & v_{1x} & v_{2x} & v_{3x}\\
      v_{0y} & v_{1y} & v_{2y} & v_{3y}\\
      v_{0z} & v_{1z} & v_{2z} & v_{3z}
  \end{matrix} \right)
  \left(\begin{matrix}
    \mu_0 \\
    \mu_1 \\
    \mu_2 \\
    \mu_3
  \end{matrix}\right) = \lambda
  \left(\begin{matrix}
    w_{0x} & w_{1x} & w_{2x} & w_{3x}\\
    w_{0y} & w_{1y} & w_{2y} & w_{3y}\\
    w_{0z} & w_{1z} & w_{2z} & w_{3z}
  \end{matrix}\right)
  \left(\begin{matrix}
    \mu_0 \\
    \mu_1 \\
    \mu_2 \\
    \mu_3
  \end{matrix}\right),
\end{equation}
where columns $(v_{ix}, v_{iy}, v_{iz})^\intercal$ and $(w_{ix}, w_{iy}, w_{iz})^\intercal$ denote the $xyz$-components of $\mathbf{v}_i$ and $\mathbf{w}_i$, respectively, on the $i$th vertex.  Because $\mu_0 + \mu_1 + \mu_2 + \mu_3=1$, we reduce $\mu_3$ and transform both sides of the equation as follows: 
\begin{align}
  & \left(\begin{matrix}
        v_{0x} - v_{3x} & v_{1x} - v_{3x} & v_{2x} - v_{3x}\\
        v_{0y} - v_{3y} & v_{1y} - v_{3y} & v_{2y} - v_{3y}\\
        v_{0z} - v_{3z} & v_{1z} - v_{3z} & v_{2z} - v_{3z}
    \end{matrix} \right)
    \left(\begin{matrix}
      \mu_0 \\
      \mu_1 \\
      \mu_2
    \end{matrix}\right) + 
    \left(\begin{matrix}
      v_{3x}\\
      v_{3y}\\
      v_{3z}
    \end{matrix}\right)\nonumber\\
= & \lambda\left[
    \left(\begin{matrix}
        w_{0x} - w_{3x} & w_{1x} - w_{3x} & w_{2x} - w_{3x}\\
        w_{0y} - w_{3y} & w_{1y} - w_{3y} & w_{2y} - w_{3y}\\
        w_{0z} - w_{3z} & w_{1z} - w_{3z} & w_{2z} - w_{3z}
    \end{matrix}\right)
    \left(\begin{matrix}
      \mu_0 \\
      \mu_1 \\
      \mu_2
    \end{matrix}\right) +
    \left(\begin{matrix}
      w_{3x}\\
      w_{3y}\\
      w_{3z}
    \end{matrix}\right)\right].\label{eq:pvtet}
\end{align}

\subsection{Generalized underdetermined eigensystem}

We define Equation~\ref{eq:pvtet} as a \emph{generalized underdetermined eigensystem} and rewrite it as 
\begin{equation}
  \mathbf{A}\bm{\mu} + \mathbf{a} = \lambda \left(\mathbf{B}\bm{\mu} + \mathbf{b}\right),
  \label{eq:underdetermined}
\end{equation}
where $\mathbf{A}$ and $\mathbf{B}$ are the $3\times3$ matrices on the left- and right-hand side of Eq.~\eqref{eq:pvtet}, respectively; $\mathbf{a}$ and $\mathbf{b}$ are equal to $\left(v_{3x}, v_{3y}, v_{3z}\right)^\intercal$ and $\left(w_{3x}, w_{3y}, w_{3z}\right)^\intercal$, respectively; and $\bm{\mu}$ represents the three independent components of the barycentric coordinates $\left(\mu_0, \mu_1, \mu_2\right)^\intercal$.\footnote{Equation~\ref{eq:underdetermined} appears similar to but is fundamentally different from those of generalized eigenvalue problems (in the form of $\mathbf{A}\bm{\mu} = \lambda\mathbf{B}\bm{\mu}$), which typically has a finite number of solutions of $\lambda$, whereas a generalized underdetermined eigensystem has infinitely many solutions of $\lambda$.}

To get the closed-form solutions of $\lambda$ and $\bm{\mu}$, we first transform Eq.~\eqref{eq:underdetermined} into
\begin{equation}
  \left(\mathbf{A} - \lambda \mathbf{B}\right)\bm{\mu} = 
  - \left(\mathbf{a} - \lambda \mathbf{b}\right).  
\end{equation}
We then left multiply the adjugate of $\left(\mathbf{A} - \lambda \mathbf{B}\right)$ on both sides of the equation, and we have
\begin{equation}
  \operatorname{adj}\left(\mathbf{A} - \lambda \mathbf{B}\right) \left(\mathbf{A} - \lambda \mathbf{B}\right)\bm{\mu} = 
  - \operatorname{adj}\left(\mathbf{A} - \lambda \mathbf{B}\right) \left(\mathbf{a} - \lambda \mathbf{b}\right), 
\end{equation}
where $\operatorname{adj}(\cdot)$ is the adjugate operator for square matrices.  Because $\operatorname{adj}(\mathbf{M})\mathbf{M}=\operatorname{det}(\mathbf{M})\mathbf{I}$ always holds for any $n\times n$ square matrix $\mathbf{M}$ even if $\mathbf{M}$ is singular, $\operatorname{det}(\mathbf{M})$ being determinant of $\mathbf{M}$ and $\mathbf{I}$ being the $n\times n$ identity matrix, we have
\begin{equation}
  \operatorname{det}\left(\mathbf{A} - \lambda \mathbf{B}\right) \bm{\mu} = - \operatorname{adj}\left(\mathbf{A} - \lambda \mathbf{B}\right) \left(\mathbf{a} - \lambda \mathbf{b}\right). \label{eq:adjugate}
\end{equation}
We will show that both $\operatorname{det}\left(\mathbf{A} - \lambda \mathbf{B}\right)$ and $\operatorname{adj}\left(\mathbf{A} - \lambda \mathbf{B}\right) \left(\mathbf{a} - \lambda \mathbf{b}\right)$ are polynomials of $\lambda$ up to degree three. Thus each component of $\bm{\mu}$ can be written as a cubic rational function of $\lambda$ when $\operatorname{det}\left(\mathbf{A} - \lambda \mathbf{B}\right)\neq \mathbf{0}$.  For simplicity, we denote $Q(\lambda)=\operatorname{det}\left(\mathbf{A} - \lambda \mathbf{B}\right)$; $P_i(\lambda)$ is the $i$th component of the 3-dimensional vector $-\operatorname{adj}\left(\mathbf{A} - \lambda \mathbf{B}\right) \left(\mathbf{a} - \lambda \mathbf{b}\right)$. We then have 
\begin{equation}
  \left\{
  \begin{array}{l}
    \mu_0 = P_0(\lambda) / Q(\lambda) \\
    \mu_1 = P_1(\lambda) / Q(\lambda) \\
    \mu_2 = P_2(\lambda) / Q(\lambda)
  \end{array}\right.\label{eq:rational}
\end{equation}
if $Q(\lambda)\neq 0$.  Because $\mu_0+\mu_1+\mu_2+\mu_3=1$, by letting $P_3(\lambda) \equiv Q(\lambda) - P_0(\lambda) - P_1(\lambda) - P_2(\lambda)$, we can also write $\mu_3$ as a rational: 
\begin{equation}
  \mu_3 = 1 - \mu_0 - \mu_1 - \mu_2 = \frac{Q(\lambda) - P_0(\lambda) - P_1(\lambda) - P_2(\lambda)}{Q(\lambda)} = \frac{P_3(\lambda)}{Q(\lambda)}.
\end{equation}
In the rest of this section, we will demonstrate that both $P_i(\lambda)$ and $Q(\lambda)$ are cubic polynomials, thus supporting our claim that PV curves are piecewise cubic rational parametric curves.

\subsection{Denominator polynomial $Q(\lambda)$}

The denominator $\operatorname{det}\left(\mathbf{A} - \lambda \mathbf{B}\right)$ is a polynomial up to the third degree: 
\begin{align}
  Q(\lambda) = & \operatorname{det}\left(\mathbf{A} - \lambda \mathbf{B}\right) = 
  \left|
  \begin{smallmatrix}
    a_{00}-b_{00}\lambda & a_{01}-b_{01}\lambda & a_{02}-b_{02}\lambda \\
    a_{10}-b_{10}\lambda & a_{11}-b_{11}\lambda & a_{12}-b_{12}\lambda \\
    a_{20}-b_{20}\lambda & a_{21}-b_{21}\lambda & a_{22}-b_{22}\lambda
  \end{smallmatrix}
  \right|\nonumber\\
  = & -\operatorname{det}(\mathbf{B})\lambda^3 + 
  \left(\left|\begin{smallmatrix}
    a_{00} & b_{01} & b_{02} \\
    a_{10} & b_{11} & b_{22} \\
    a_{20} & b_{21} & b_{22}
  \end{smallmatrix}\right| + 
  \left|\begin{smallmatrix}
    b_{00} & a_{01} & b_{02} \\
    b_{10} & a_{11} & b_{22} \\
    b_{20} & a_{21} & b_{22}
  \end{smallmatrix}\right| + 
  \left|\begin{smallmatrix}
    b_{00} & b_{01} & a_{02} \\
    b_{10} & b_{11} & a_{22} \\
    b_{20} & b_{21} & a_{22}
  \end{smallmatrix}\right|\right)\lambda^2\nonumber\\
  & - 
  \left(\left|\begin{smallmatrix}
    a_{00} & a_{01} & b_{02} \\
    a_{10} & a_{11} & b_{22} \\
    a_{20} & a_{21} & b_{22}
  \end{smallmatrix}\right| + 
  \left|\begin{smallmatrix}
    a_{00} & b_{01} & a_{02} \\
    a_{10} & b_{11} & a_{22} \\
    a_{20} & b_{21} & a_{22}
  \end{smallmatrix}\right| + 
  \left|\begin{smallmatrix}
    b_{00} & a_{01} & a_{02} \\
    b_{10} & a_{11} & a_{22} \\
    b_{20} & a_{21} & a_{22}
  \end{smallmatrix}\right|\right)\lambda +
   \operatorname{det}(\mathbf{A}),
\end{align}
where $a_{ij}$ and $b_{ij}$, respectively, is the $i$th row and $j$th column of $\mathbf{A}$ and $\mathbf{B}$.  We will refer to $Q(\lambda)$ as the \emph{characteristic polynomial} of the given tetrahedron.  

In general, the roots of the third-degree polynomial $Q(\lambda)$ can be written in closed form.  In special cases when the cubic coefficient $\operatorname{det}(\mathbf{B})$ is zero (or the quadratic coefficient coincidentally being zero too), the roots are still in closed form. 


\subsection{Numerator polynomials $P_i(\lambda)$}
\label{sec:numerator}

Each component of the right-hand side of Equation~\eqref{eq:adjugate}, that is, $-\operatorname{adj}\left(\mathbf{A} - \lambda \mathbf{B}\right) \left(\mathbf{a} - \lambda \mathbf{b}\right)$, is a polynomial of $\lambda$ up to the third degree.  The first part of the product is a $3\times3$ adjugate matrix:
\begin{align}
& \operatorname{adj}\left(\mathbf{A} - \lambda \mathbf{B}\right)\nonumber \\
= & 
\left(
\begin{smallmatrix}
  + \left|\begin{smallmatrix}
      a_{11} - b_{11}\lambda & a_{12} - b_{12}\lambda \\
      a_{21} - b_{21}\lambda & a_{22} - b_{22}\lambda
    \end{smallmatrix}\right| & 
  - \left|\begin{smallmatrix}
      a_{01} - b_{01}\lambda & a_{02} - b_{02}\lambda \\
      a_{21} - b_{21}\lambda & a_{22} - b_{22}\lambda
    \end{smallmatrix}\right|
  + \left|\begin{smallmatrix}
      a_{01} - b_{01}\lambda & a_{02} - b_{02}\lambda \\
      a_{11} - b_{11}\lambda & a_{12} - b_{12}\lambda
    \end{smallmatrix}\right|\\
  - \left|\begin{smallmatrix}
      a_{10} - b_{10}\lambda & a_{12} - b_{12}\lambda \\
      a_{20} - b_{20}\lambda & a_{22} - b_{22}\lambda
    \end{smallmatrix}\right| & 
  + \left|\begin{smallmatrix}
      a_{00} - b_{00}\lambda & a_{02} - b_{02}\lambda \\
      a_{20} - b_{20}\lambda & a_{22} - b_{22}\lambda
    \end{smallmatrix}\right|
  - \left|\begin{smallmatrix}
      a_{00} - b_{00}\lambda & a_{02} - b_{02}\lambda \\
      a_{10} - b_{10}\lambda & a_{12} - b_{12}\lambda
    \end{smallmatrix}\right|\\
  + \left|\begin{smallmatrix}
      a_{10} - b_{10}\lambda & a_{11} - b_{11}\lambda \\
      a_{20} - b_{20}\lambda & a_{21} - b_{21}\lambda
    \end{smallmatrix}\right| & 
  - \left|\begin{smallmatrix}
      a_{00} - b_{00}\lambda & a_{01} - b_{11}\lambda \\
      a_{20} - b_{20}\lambda & a_{21} - b_{21}\lambda
    \end{smallmatrix}\right|
  + \left|\begin{smallmatrix}
      a_{00} - b_{00}\lambda & a_{01} - b_{01}\lambda \\
      a_{10} - b_{10}\lambda & a_{11} - b_{11}\lambda
    \end{smallmatrix}\right|
\end{smallmatrix}
\right);
\end{align}
each element is a $2\times 2$ determinant, which is a polynomial of $\lambda$ up to degree two.  The second part of the product is the $3\times 1$ vector $\left(\mathbf{a} - \lambda \mathbf{b}\right)$; each component is a degree-one polynomial of $\lambda$.  Each component of the product $- \operatorname{adj}\left(\mathbf{A} - \lambda \mathbf{B}\right) \left(\mathbf{a} - \lambda \mathbf{b}\right)$ is thus a polynomial up to degree three, denoted as $P_i(\lambda)$.  

A key observation can be made by studying the full expansion\footnote{We omit the very long expansion for space.} of $-\operatorname{adj}\left(\mathbf{A} - \lambda \mathbf{B}\right) \left(\mathbf{a} - \lambda \mathbf{b}\right)$: coefficients of $P_i(\lambda)$  contain only values of $\mathbf{v}_j$ and $\mathbf{w}_j$, $j\in\{0, 1, 2, 3\}$ and $j\neq i$.  
For example, $P_3(\lambda)$ is  related only to the values of $\mathbf{v}_0$, $\mathbf{w}_0$, $\mathbf{v}_1$, $\mathbf{w}_1$, $\mathbf{v}_2$, and $\mathbf{w}_2$.  In this case, $P_3(\lambda)$ is the characteristic polynomial of Equation~\eqref{eq:eig3}, which is the basis of the Peikert--Roth  method~\cite{PeikertR99} for extracting parallel vector points on triangular faces.

\subsection{Degeneracies}

We discuss degeneracy cases when $Q(\lambda)$ becomes zero.  

First, if $Q(\lambda)$ constantly equals $0$, there is no solution to the equation if $P_i(\lambda)\neq0$ for all $i$ unless the right-hand side of Equation~\ref{eq:adjugate} is $\mathbf{0}$.  An example of the latter case ($P_i(\lambda)=Q(\lambda)=0$ for all $i$ and $\lambda$) is $\mathbf{v}(\mathbf{x})=\mathbf{w}(\mathbf{x})=\mathbf{0}$, which satisfy $\mathbf{v}(\mathbf{x})\times\mathbf{w}(\mathbf{x})=\mathbf{0}$ everywhere in the tetrahedron; we do not consider such degeneracy cases in PV curve extraction.

Second, if $Q(\lambda)$ has a real root $\lambda_0$, it is typically a degeneracy case unless the limit $\lim_{\lambda\to\lambda_0}P_i(\lambda)/Q(\lambda)$ exists for all $i\in\{0, 1, 2, 3\}$.  If the limit exists, $\lambda_0$ is a common root of $P_i(\lambda)$ and $Q(\lambda)$.


%


\section{Parallel Vector Curves inside a Tetrahedron}
\label{sec:tetrahedron}

The extraction of parallel vector curves inside a tetrahedron is equivalent to the solutions of $\mu_0$, $\mu_1$, $\mu_2$, and $\mu_3\in[0,1]$, that is,
\begin{equation}
  \left\{
  \begin{array}{l}
    0 \leq P_0(\lambda) / Q(\lambda) \leq 1\\
    0 \leq P_1(\lambda) / Q(\lambda) \leq 1\\
    0 \leq P_2(\lambda) / Q(\lambda) \leq 1\\
    0 \leq P_3(\lambda) / Q(\lambda) \leq 1
  \end{array}\right.,\label{eq:inequalities}
\end{equation}
which further leads to following eight distinct inequalities.
\begin{equation}\label{eq:inequalities}
  \begin{array}{cc}
    P_0(\lambda) / Q(\lambda) \geq 0 & (Q(\lambda) - P_0(\lambda)) / Q(\lambda) \geq 0\\
    P_1(\lambda) / Q(\lambda) \geq 0 & (Q(\lambda) - P_1(\lambda)) / Q(\lambda) \geq 0\\
    P_2(\lambda) / Q(\lambda) \geq 0 & (Q(\lambda) - P_2(\lambda)) / Q(\lambda) \geq 0\\
    P_3(\lambda) / Q(\lambda) \geq 0 & (Q(\lambda) - P_3(\lambda)) / Q(\lambda) \geq 0
  \end{array}.
\end{equation}
One can find a finite number of intervals of $\lambda\in\mathbb{R}$; each interval corresponds to a disjoint branch of the curve in the tetrahedron.

\subsection{Solution intervals of each cubic rational inequality}
\label{sec:solution}

Without loss of generality, let $P(\lambda)/Q(\lambda)\geq 0$ be any of the inequalities in Equation~\eqref{eq:inequalities}; $P(\lambda)$ and $Q(\lambda)$ are cubic polynomials.  We describe the method by assuming that $P(\lambda)$ and $Q(\lambda)$ do not share any roots; if there exists $q$ such that $P(q)=Q(q)=0$, we first reduce the rational $\frac{P(\lambda)}{Q(\lambda)}$ to $\frac{P(\lambda)}{\lambda-q}/\frac{Q(\lambda)}{\lambda-q}$ and then use the new numerator and denominator as the input to solve the inequality.  

We solve the inequality by (1) finding all roots of $P(\lambda)$ and $Q(\lambda)$, (2) sorting the roots such that $-\infty < r_0 < r_1 < \ldots < r_{n_r} < +\infty$, $n_r$ being the total number of roots, and (3) checking whether  $P(\lambda)/Q(\lambda)\geq0$ for each interval $(-\infty, r_0), (r_1, r_2), \ldots, (r_{n_r}, +\infty)$.  Note that each endpoint of the result intervals is open if the endpoint is a root of $Q(\lambda)$; otherwise the endpoint is closed.  For example, the result may be $(-\infty, r_0)\cup[r_2, +\infty)$ if $Q(r_0)=0$ and $Q(r_2)\neq 0$; another result may be $[r_2, r_3]$ if none of $r_2$ and $r_3$ is the root of $Q(\lambda)$.  We will interpret the meaning of open and closed intervals in the next subsection.

\begin{figure}[h]
  \includegraphics[width=\linewidth]{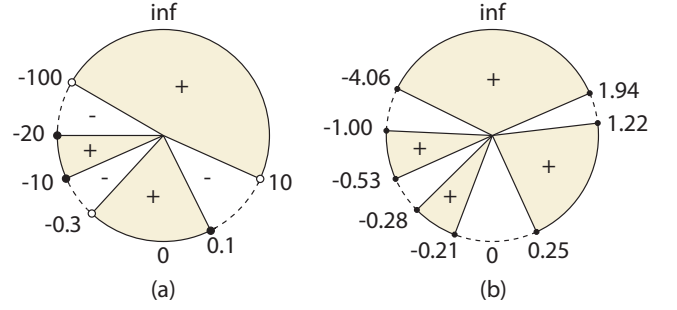}
  \caption{Ring representation of two sets of rational inequality solutions.  Solid dots and hollow dots, respectively, are the solutions of the numerators and denominators.}
  \label{fig:ring}
\end{figure}

\begin{figure*}[h]
  \includegraphics[width=\linewidth]{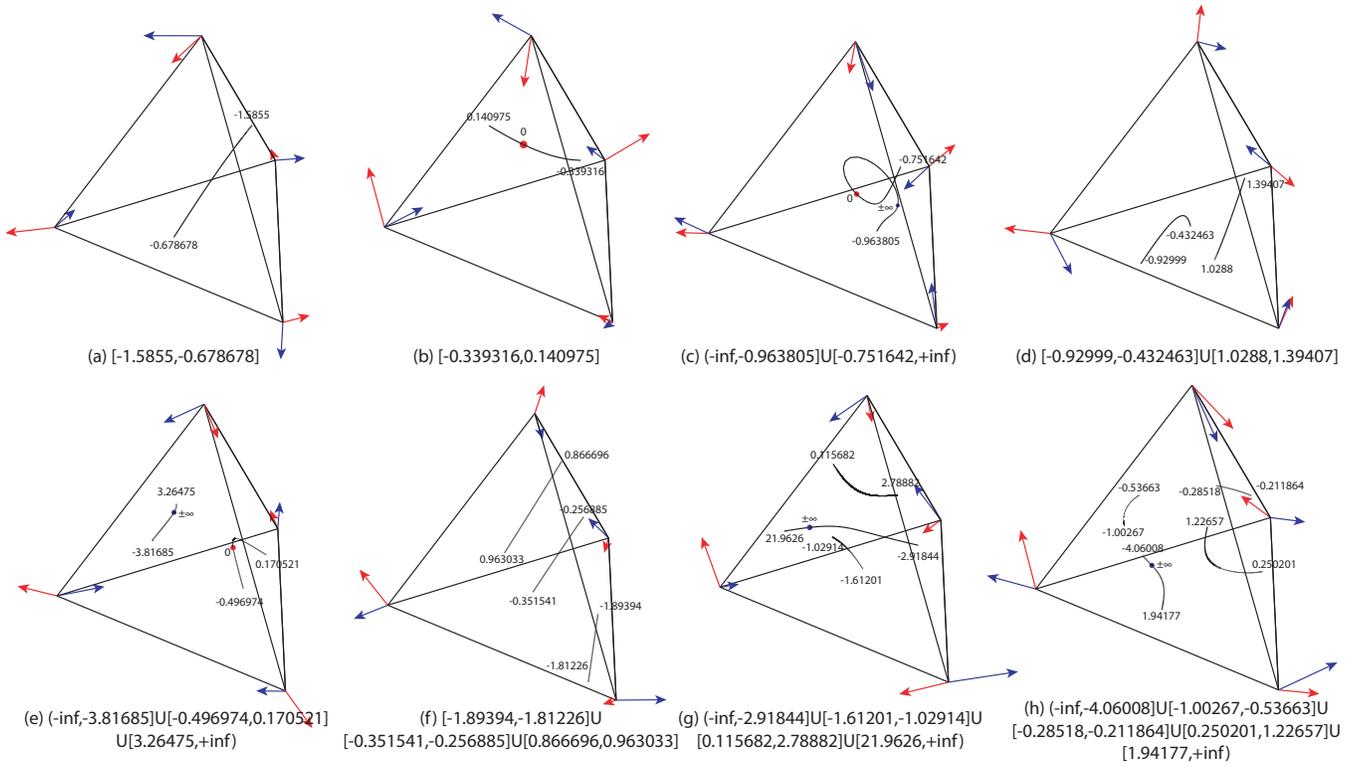}
  \caption{Possible configurations of PV curves intersecting a tetrahedron: (a) single branch with a normal interval, (b) single branch with a critical point of $\mathbf{v}$, (c) single branch with a critical point of $\mathbf{v}$ and $\mathbf{w}$, respectively, (d) two branches with normal intervals, (e) two branches with a critical point on each branch, (f) three branches, (g) three branches with a critical point of $\mathbf{w}$ on one of the branches, (h) four branches with a critical point of $\mathbf{w}$ on one of the branches.  Red and blue arrows, respectively, indicate the $\mathbf{v}$ and $\mathbf{w}$ vectors.}\label{fig:cases}
\end{figure*}

We consider $\lambda$ in the extended real domain $\bar{\mathbb{R}} = \mathbb{R}\cup\{\infty\}$.  This is reasonable because normally (when $P(\lambda)$ and $Q(\lambda)$ have nonzero cubic coefficients) the limits of $P(\lambda)/Q(\lambda)$ at positive and negative infinities exist and are equal:
\begin{equation}
  \lim_{\lambda\to\infty} \frac{P(\lambda)}{Q(\lambda)} = \lim_{\lambda\to-\infty} \frac{P(\lambda)}{Q(\lambda)} = \lim_{\lambda\to+\infty} \frac{P(\lambda)}{Q(\lambda)}.
\end{equation}
Thus, we view two intervals that share the infinity endpoint as one single interval.  For example, $(-\infty, 10]\cup[20, +\infty)$ is considered as one single interval that contains the infinity.  

Figure~\ref{fig:ring} illustrates $\bar{\mathbb{R}}$ as a ring; the bottom and top of the ring are zero and infinity, respectively.  Solutions of $P(\lambda)$ and $Q(\lambda)$, respectively, are mapped to solid and hollow dots on the ring.  The color of each sector indicates the sign of $P(\lambda)/Q(\lambda)$.  Figure~\ref{fig:ring}(a) illustrates the solution of $P(\lambda)=(\lambda+20)(\lambda+10)(\lambda-0.1)$ and $Q(\lambda)=(\lambda+100)(\lambda+0.3)(\lambda-10)$; as a result, the solutions of $P(\lambda)/Q(\lambda)\geq0$ are $(-\infty, -100)\cup[-20, 10]\cup(-0.3,0.1]\cup(10, +\infty)$.  

%


\subsection{Solutions of all rational inequalities}

The solution of Equation~\eqref{eq:inequalities} is the intersections of solutions of individual inequalities.  As a result, the solution is either an empty set or the union of subintervals. 
If the result is an empty set, the parallel vector curve does not intersect the tetrahedra; 
otherwise there exists intersections.  The number of subintervals ranges from zero to four, and each interval corresponds to a continuous segment of the parallel vector curve.  Normally, each subinterval is a closed interval because the \emph{feasible region} of $\lambda$ such that $P_i(\lambda)/Q(\lambda)\in[0, 1]$ is closed.  For example, Figure~\ref{fig:ring}(b) illustrates solutions of multiple rational inequalities, leading to four feasible regions.

Figure~\ref{fig:cases} demonstrates possible configurations that PV curve intersect a tetrahedron with synthetic data.  If infinity is included in the interval, a \emph{critical point} exists in $\mathbf{w}(\mathbf{x})$ at the location where $\lambda$ is infinity.  The interval of $(-\infty, \lambda_0]\cup[\lambda_1, +\infty)$. 
As we increase (or decrease) $\lambda$ from $\lambda_1$ (or $\lambda_0$) to $+\infty$ (or $-\infty$), the PV point converges to the critical point where $\mathbf{w}(\mathbf{x})=\mathbf{0}$.

%% file: algorithm.tex
\section{PV curve reconstruction}
\label{sec:alg}

We present a two-pass algorithm to reconstruct PV curves: the first pass computes and solves the numerator polynomial for each triangular face in the mesh, and the second pass computes the denominator polynomial and extracts PV curves inside each tetrahedron.

\textbf{Per-triangle numerator pass.}\quad
We calculate the exact roots of the numerator polynomial ($P(\lambda)$) for each triangular face.  The cubic numerator polynomial typically has up to three real roots, each corresponding to an intersection between a PV curve and the plane that contains the triangle.  If the intersection is inside the triangle, we record the tuple of triangle ID, $\lambda$ value, and the barycentric coordinates of the intersection for the next pass.
%

\textbf{Per-tetrahedron denominator pass.}\quad
We compute the denominator polynomial ($Q(\lambda)$) and its root(s) for every tetrahedra that are labeled in the previous pass, in order to reconstruct PV curves.  We gather the roots of all numerator polynomials of the triangular sides and then solve the solution intervals.  As a result, each tetrahedron in the iteration finds one or multiple closed intervals of $\lambda$, and each interval corresponds to a segment of PV curves.

\begin{figure}
  \includegraphics[width=\linewidth]{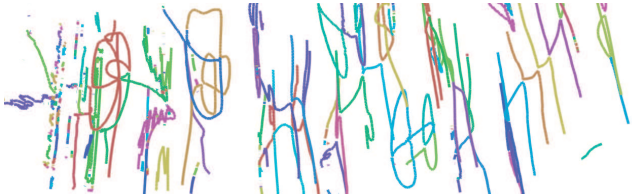}
  \caption{PV curves extracted from a flow-past-cylinder dataset.}
  \label{fig:cylinder}
\end{figure}


\textbf{Results and limitations.}\quad Figure~\ref{fig:cylinder} shows PV curves extracted from a flow-past-cylinder dataset.  We acknowledge the limitation of the piecewise linearity assumption on both $\mathbf{v}$ and $\mathbf{w}$.  First, one has to tessellate the input mesh if the input data are not given in tetrahedral mesh.  For example, if the input data are in a regular grid, one may subdivide each cube into a number of tetrahedra; however, multiple possible subdivisions exist and may lead to different PV extraction results.  Second, one has to make assumptions about the linearity.  For example, the Sujudi--Haimes descriptor, $\mathbf{w}=(\nabla\mathbf{v})\mathbf{v}$ is not linear even if $\mathbf{v}$ is linear; in future work, we will investigate the error of PV curves when $\mathbf{w}$ is interpolated linearly.

%% file: conclusions.tex
\section{Conclusions}
\label{sec:conclusions}

This paper proves that PV curves are cubic rational curves in two linear vector fields and presents an analytical exact PV curve extraction algorithm.  We believe this work opens numerous research avenues.  First, one can develop methods to query, filter, and simplify PV curves for feature exploration.  Second, one can investigate the change of $\lambda$ values along PV curves; although $\lambda$ is monotonous within each tetrahedron, the change over the entire curve may reveal key insights into the data.  Third, one  can build connections between vector field topology (e.g., critical points) with PV curves.  Fourth, it would be straightforward to further generalize the derivation to track PV curves over time, in order to capture the dynamics of key features in time-varying scientific data.  Fifth, the two-pass reconstruction algorithm can be directly accelerated with both GPUs and distributed parallel  computing for analyzing very large data.